\documentclass[aoas]{imsart}

\usepackage{courier}  
\usepackage[hyphens]{url}  
\usepackage{graphicx} 
\usepackage{hyperref} 

\usepackage{caption} 
\usepackage{algorithm}

\usepackage{amsmath}
\usepackage{amssymb}
\usepackage{algorithm}
\usepackage[noend]{algpseudocode}

\algnewcommand\algorithmicinput{\textbf{Input:}}
\algnewcommand\Input{\item[\algorithmicinput]}
\algnewcommand\algorithmicoutput{\textbf{Output:}}
\algnewcommand\Output{\item[\algorithmicoutput]}

\usepackage{booktabs} 
\usepackage{amsmath}  
\usepackage{multirow}

\usepackage{newfloat}
\usepackage{listings}

\newcommand\blfootnote[1]{%
  \begingroup
  \renewcommand\thefootnote{}\footnote{#1}%
  \addtocounter{footnote}{-1}%
  \endgroup
}

\RequirePackage{amsthm,amsmath,amsfonts,amssymb}
\RequirePackage[authoryear]{natbib}
\startlocaldefs

\endlocaldefs

\begin{document}

\begin{frontmatter}
\title{Accept-Reject Lasso}

\begin{aug}
\author[A]{\fnms{Yanxin}~\snm{Liu}\ead[label=e1]{12024213042@stu.ynu.edu.cn}},
\author[A]{\fnms{Yunqi}~\snm{Zhang}\ead[label=e2]{yunqizhang@ynu.edu.cn}}

\address[A]{Yunnan University\printead[presep={,\ }]{e1}}

\address[A]{Yunnan University\printead[presep={,\ }]{e2}}
\end{aug}

\begin{abstract}
The Lasso method is known to exhibit instability in the presence of highly correlated features, often leading to an arbitrary selection of predictors. This issue manifests itself in two primary error types: the erroneous omission of features that lack a true substitutable relationship (falsely redundant features) and the inclusion of features with a true substitutable relationship (truly redundant features). Although most existing methods address only one of these challenges, we introduce the Accept-Reject Lasso (ARL), a novel approach that resolves this dilemma. ARL operationalizes an Accept-Reject framework through a fine-grained analysis of feature selection across data subsets. This framework is designed to partition the output of an ensemble method into beneficial and detrimental components through fine-grained analysis. The fundamental challenge for Lasso is that inter-variable correlation obscures the true sources of information. ARL tackles this by first using clustering to identify distinct subset structures within the data. It then analyzes Lasso's behavior across these subsets to differentiate between true and spurious correlations. For truly correlated features, which induce multicollinearity, ARL tends to select a single representative feature and reject the rest to ensure model stability. Conversely, for features linked by spurious correlations, which may vanish in certain subsets, ARL accepts those that Lasso might have incorrectly omitted. The distinct patterns arising from true versus spurious correlations create a divisible separation. By setting an appropriate threshold, our framework can effectively distinguish between these two phenomena, thereby maximizing the inclusion of informative variables while minimizing the introduction of detrimental ones. We illustrate the efficacy of the proposed method through extensive simulation and real-data experiments.
\end{abstract}


\end{frontmatter}

\section{Introduction}

Lasso regression \cite{1}, valued for its inherent features selection capabilities and the interpretability of its models, remains a widely used technique in statistical learning. However, a well-known issue with the standard Lasso is its behavior when faced with a group of highly correlated features: it tends to select only one or a few of these features, while shrinking the coefficients of the others to exactly zero.

Although selecting a single representative feature from a correlated group can be an effective strategy to mitigate multicollinearity, this approach becomes problematic if the observed correlation is spurious. Spurious correlations may arise from data collection biases rather than reflecting a true underlying relationship. In such scenarios, omitting these falsely correlated variables can lead to a significant degradation in the model's predictive accuracy and explanatory power when deployed in real-world applications.

To formalize this problem, consider the linear model:
$$
y = \sum_{k=1}^{p} \beta_{k} x_{k}
$$
Let $\mathcal{P}$ be the set of all features. We posit that within the subset of significant features in $\mathcal{P}$, two distinct types of correlation structures can exist.
First, there are groups of features that exhibit a \textbf{True Redundancy (TR)}. The features within a TR group are genuinely substitutable for one another. Including all features from a TR group would introduce severe multicollinearity \cite{8,9,10}, thereby impairing model interpretability and predictive performance. Consequently, selecting a single representative feature from each TR group is the optimal strategy.
Second, there are groups of features characterized by a \textbf{False Redundancy (FR)}. The correlation among features in an FR group is an artifact of the specific dataset, for instance, due to sampling bias, and does not hold universally. These features are not genuinely substitutable.

This dichotomy is common in practice. A classic example is Simpson's paradox \cite{11}, which can be observed in medical studies. Consider two hospitals: Hospital A, which specializes in treating mild cases, and Hospital B, which treats severe cases. A new drug might be ineffective in both hospitals. However, because patients in Hospital B have an inherently higher mortality rate (regardless of treatment) and the new drug is administered more frequently at Hospital B, a pooled analysis of the data could create a spurious correlation between the use of the new drug and a higher mortality rate. This illustrates the precise dilemma that standard feature selection methods face.

In such contexts, an ideal feature selection method should retain all variables from the FR groups while selecting only one or a few representative variables from each TR group. We define two types of Lasso errors. A \textbf{Type I Lasso Error} occurs when (1) an FR feature is incorrectly omitted, or (2) all features from a TR group are discarded due to extreme correlation. A \textbf{Type II Lasso Error} occurs when too many redundant features from a TR group are selected. Existing methods often struggle to control both types of errors simultaneously.

To address this challenge, we focus on subsets of features exhibiting high internal correlation, which we term \textbf{problem groups}. We denote the $q$-th problem group as $Q_q = \{p_a\}$, where all features within $Q_q$ are highly correlated with each other. Here, $q \in \{1, \dots, r\}$, where $r$ is the total number of problem groups.

\section{Related Work}
Existing enhancements to Lasso do not adequately resolve this problem. For instance, among non-ensemble methods, the Adaptive Lasso \cite{2} assigns smaller penalty weights to more important features and larger weights to less important ones; the SLOPE penalty \cite{12,13}, in contrast, posits that more significant features should receive larger penalties to control the False Discovery Rate (FDR); the Elastic Net \cite{3} incorporates an L2-norm penalty to mitigate Lasso's tendency to drop correlated variables; and the Group Lasso \cite{7}, which presupposes a group structure among features, allows for the simultaneous inclusion or exclusion of entire pre-defined groups by applying penalties at the group level. However, none of these methods possesses a mechanism to differentiate between true and false redundancy.

Ensemble methods, on the other hand, typically excel at controlling only one type of error. For example, \textbf{Stability Selection} \cite{5} uses bootstrapping \cite{11} to generate multiple data sets. By aggregating the feature selection results from running Lasso on each bootstrap sample, it aims to produce a more robust final feature set. This process is effective at eliminating noise and TR features, thereby controlling the Type II Lasso Error. However, the bootstrap samples are drawn from the overall population and thus tend to replicate its correlation structure. The core principle of stability selection—retaining frequently selected variables—is fundamentally at odds with Lasso's tendency to randomly select one feature from a highly correlated group. This can lead to a situation where different features from the same group are chosen in different bootstrap samples, causing each feature's selection frequency to fall below the retention threshold. As a result, the Type I Lasso Error is not effectively controlled.

\blfootnote{The implementation of the ARL algorithm and the complete pipeline to reproduce all experimental results in this paper are publicly available at \url{https://github.com/liudaohe/Accept-Reject-Lasso.git}}

In contrast, the \textbf{Random Lasso} \cite{4} also leverages bootstrapping, but averages the coefficients of the selected variables across all samples. This approach is designed to mitigate Lasso's tendency to omit variables, thus addressing the Type I Lasso Error. However, by preserving features that are selected even in a few bootstrap samples, it is prone to including multiple, truly redundant variables from TR groups, failing to control the Type II Lasso Error.

In summary, current algorithms cannot concurrently retain features from FR groups while correctly identifying and consolidating TR groups, i.e., they cannot control both types of Lasso errors in the scenarios we consider.

\section{Accept-Reject Lasso}

To address this gap, we propose a novel algorithm, the \textbf{Accept-Reject Lasso (ARL)}. Our approach is inspired by the Accept-Reject framework \cite{6}, which posits that an uncontrollable ensemble learning process, reliant on expectations over a large sample, can be transformed into a controllable one through finer-grained partitioning. This allows for the separation of beneficial and detrimental variations within the ensemble, such that "good" changes are accepted and "bad" ones are rejected.

In our context, the primary issue with the aforementioned ensemble methods is their inability to separate the "good" improvements from the "bad." Stability Selection, in controlling Type II errors, inadvertently introduces Type I errors, while Random Lasso, in controlling Type I errors, introduces Type II errors. We need an ensemble that can distinguish between true and false redundancy, allowing us to rescue erroneously omitted FR features without incorrectly including multiple TR features. Through this mechanism, we can \textbf{accept} the inclusion of falsely redundant features while \textbf{rejecting} the superfluous ones from TR groups.

Our core hypothesis is as follows: a true correlation should be stable and manifest across any representative subset of the data. Conversely, a spurious correlation may be an artifact of the global dataset and might break down within specific, more homogeneous subsets.

Based on this principle, the ARL algorithm partitions the global dataset into several locally distinct subsets via clustering. By running Lasso independently on these subsets, we can examine the stability of the correlation structures. If a group of features is TR, Lasso should consistently select only one or a few of them in every subset. If the redundancy is FR, the features contain independent information that may lead Lasso to select all of them simultaneously in certain subsets.

By analyzing the co-occurrence frequency of features within the problem groups across these subsets, we can diagnose FR. If a group of features frequently co-occurs in the selection results of the subsets, we infer the presence of an FR group and rescue all of its features.

The specific steps of the ARL algorithm are as follows, with the full dataset denoted as $D$:

\begin{enumerate}
    \item \textbf{Initial Feature Selection:} Run a baseline Lasso on the entire dataset $D$ to obtain an initial set of selected features, denoted as $P_G$.

    \item \textbf{Identify Problem Groups:} As previously discussed, Lasso errors occur within groups of highly correlated variables. We construct a feature correlation matrix and model the features as nodes in a graph. An edge is placed between any two features if the absolute value of their correlation exceeds a threshold $\tau_{corr}$. The connected components of this graph constitute the problem groups. Let the set of all problem groups be $\mathcal{Q} = \{Q_1, Q_2, \dots, Q_r\}$, where each $Q_i$ is a problem group and $Q_i \cap Q_j = \emptyset$ for $i \neq j$.
    
    A filtering mechanism can be optionally applied here. We posit that it is highly improbable for two pure noise variables to exhibit extreme correlation. To safeguard against such a scenario, we can filter the identified problem groups, retaining only those that contain at least one feature from the initial selection $P_G$ (i.e., retain $Q_i$ only if $Q_i \cap P_G \neq \emptyset$). This step helps save a small amount of computational resources and prevents noise variables from being incorrectly selected later. However, this filter is often omitted because a TR group with extremely high internal correlation might have no features selected in the initial Lasso run, leading to its erroneous exclusion and thus causing a Type I Lasso Error.

    \item \textbf{Heuristic for Clustering Basis:} To effectively partition the dataset and highlight different correlation behaviors, we employ a heuristic to select a basis of features for clustering. We iterate through each problem group $Q_i$. For each $Q_i$, we perform k-means clustering on the samples using only the features in $Q_i$, setting the number of clusters to $k = |Q_i|$. The rationale is to give each feature in the group an opportunity to define its distinct sample cluster. We then compute the silhouette coefficient for this clustering. All problem groups $Q_i$ that yield a silhouette coefficient greater than a threshold $\tau_{sil}$ are considered good bases for partitioning the data. The union of all features from these selected groups forms the final clustering basis, denoted as $K$. This procedure aims to find a feature subspace $K$ that maximizes the behavioral heterogeneity between the resulting subsets.

    \item \textbf{Data Partitioning:} Perform k-means clustering on the samples using the features in the basis set $K$. The number of clusters, $m$, is a hyperparameter. Each resulting cluster of samples forms a subset, thereby partitioning the original dataset $D$ into $m$ disjoint subsets: $\{D_1, D_2, \dots, D_m\}$.

    \item \textbf{Subset Lasso:} Run the baseline Lasso independently on each subset $D_d$ using only the features from problem groups and initially selected features ($\bigcup_{Q_i \in \mathbb{Q}} Q_i \cup P_G$) to obtain a set of selected features $S_d$ for each subset.

    \item \textbf{Feature Rescue:} We rescue features based on their co-occurrence in the subset Lasso results. For each problem group $Q_i$, we consider all its subsets of size two or greater as candidate sets. For each candidate set $C$ (where $C \subseteq Q_i$ and $|C| \ge 2$), we calculate its co-occurrence frequency, defined as the number of subsets in which $C$ was entirely selected by Lasso.
    $$
    \text{count}(C) = \sum_{d=1}^{m} \mathbb{I}(C \subseteq S_d)
    $$
    where $\mathbb{I}(\cdot)$ is the indicator function.
    
    We define an integer hyperparameter, the co-occurrence threshold $\tau_{co}$. If a candidate set $C$ has a co-occurrence frequency $\text{count}(C) > \tau_{co}$, we conclude that all features within $C$ should be rescued. The final set of selected features, $P_{final}$, is the union of the initially selected features $P_G$ and all rescued features.
\end{enumerate}

The entire algorithm involves four key hyperparameters: the correlation threshold for problem groups $\tau_{corr}$, the silhouette coefficient threshold $\tau_{sil}$, the number of clusters for partitioning $m$, and the co-occurrence threshold for feature rescue $\tau_{co}$. Default values such as $\tau_{corr}=0.8$ and $\tau_{sil}=0.5$ often work well. The selection of $m$ and $\tau_{co}$ will be discussed in detail in the simulation experiments.

\begin{algorithm}
\caption{Feature Rescue via Subgroup Analysis}
\label{alg:feature_rescue}
\begin{algorithmic}[1]
\Input Dataset $D = (X, y)$; Hyperparameters: $\tau_{\text{corr}}$, $\tau_{\text{sil}}$, $m$, $\tau_{\text{co}}$.
\Output Final selected feature set $P_{\text{final}}$.

\State $P_G \gets \text{Lasso}(X, y)$ \Comment{Initial global feature selection}
\State $\mathbb{Q} \gets \text{IdentifyProblemGroups}(X, \tau_{\text{corr}})$ \Comment{Identify all problem groups}

\State $K \gets \emptyset$ \Comment{Find basis features for clustering}
\ForAll{$Q_i \in \mathbb{Q}$}
    \If{$\text{SilhouetteScore}(\text{KMeans}(X_{Q_i}, k=|Q_i|)) > \tau_{\text{sil}}$}
        \State $K \gets K \cup Q_i$
    \EndIf
\EndFor

\State $\{D_1, \dots, D_m\} \gets \text{KMeans}(X_K, k=m)$ \Comment{Partition data into $m$ subsets}

\ForAll{$d \in \{1, \dots, m\}$} \Comment{Run Lasso on each subset}
    \State $S_d \gets \text{Lasso}(X_{D_d}, y_{D_d})$ \Comment{Run Lasso on subset $D_d$ using the feature set $\bigcup_{Q_i \in \mathbb{Q}} Q_i \cup P_G$}
\EndFor

\State $P_{\text{rescued}} \gets \emptyset$ \Comment{Rescue features based on co-occurrence}
\ForAll{$Q_i \in \mathbb{Q}$}
    \ForAll{$C \in \{C' \mid C' \subseteq Q_i, |C'| \ge 2\}$}
        \If{$\sum_{d=1}^{m} \mathbb{I}(C \subseteq S_d) > \tau_{\text{co}}$}
            \State $P_{\text{rescued}} \gets P_{\text{rescued}} \cup C$
        \EndIf
    \EndFor
\EndFor

\State $P_{\text{final}} \gets P_G \cup P_{\text{rescued}}$ \Comment{Combine results}
\State \Return $P_{\text{final}}$
\end{algorithmic}
\end{algorithm}

\section{Analysis}

\subsection{Computational Complexity Analysis}

In this section, we provide a comprehensive analysis of the computational complexity of the ARL algorithm and compare it with existing methods. We begin by establishing notation, then analyze each component's complexity, and finally compare ARL with Random Lasso and Stability Selection.

\textbf{Notation and Definitions.} Let us define the following notation for our complexity analysis:

\begin{itemize}
    \item $n$: number of samples
    \item $p$: total number of features
    \item $G$: number of problem groups identified
    \item $g$: average size of problem groups
    \item $m$: number of data subsets for clustering
    \item $f$: total number of features used in subset analysis ($f = |\bigcup_{Q_i \in \mathbb{Q}} Q_i \cup P_G|$)
    \item $b$: number of basis features used for data partitioning
    \item $B$: number of bootstrap/subsampling iterations (for Random Lasso and Stability Selection)
    \item $\mathcal{C}_{\text{base}}(n, p)$: computational complexity of the baseline Lasso method
    \item $\tau_{\text{co}}$: co-occurrence threshold parameter
\end{itemize}

Note that in practice, we have $G \ll p$, $g \ll p$, $f \ll p$, and $b \ll p$ due to the sparse nature of problem groups and the effectiveness of our optimization strategies.

\textbf{Step-by-Step Complexity Analysis.} We analyze the computational complexity of each step in the ARL algorithm:

\begin{enumerate}
    \item \textbf{Global Feature Selection:} The complexity is $\mathcal{C}_{\text{base}}(n, p)$, representing the computational cost of the baseline Lasso method. For standard coordinate descent Lasso with maximum iterations $k$, this becomes $O(npk)$ as a special case.
    
    \item \textbf{Problem Group Identification:} Computing the correlation matrix requires $O(p^2)$ operations, followed by breadth-first search for connected components, which is $O(p + E)$ where $E$ is the number of edges. The total complexity is $O(p^2)$.
    
    \item \textbf{Clustering Analysis:} For each of the $G$ problem groups, we perform K-means clustering with $g$ features and $c$ iterations, resulting in $O(G \cdot n \cdot g \cdot c)$ complexity. Since $G \cdot g \ll p$ in practice, this cost is dominated by and can be absorbed into the baseline Lasso complexity.
    
    \item \textbf{Data Partitioning:} K-means clustering on $n$ samples using $b$ basis features with $c$ iterations has complexity $O(n \cdot b \cdot c)$. Since $b \ll p$, this is also dominated by the baseline Lasso complexity.
    
    \item \textbf{Subset Lasso Analysis:} We run the baseline Lasso method on $m$ subsets, each with approximately $n/m$ samples and $f$ features, where $f$ represents the union of problem group features and initially selected features. The total complexity is $m \cdot \mathcal{C}_{\text{base}}(n/m, f)$. Since $f \ll p$, this complexity is also dominated by the global baseline Lasso.
    
    \item \textbf{Co-occurrence Analysis:} For each problem group $Q_i$ and each subset, we enumerate all possible feature combinations of size 2 or larger. With our optimization that pre-filters frequent features, the complexity becomes $O(G \cdot m \cdot 2^{g})$.
\end{enumerate}

\textbf{Overall Complexity.} Since steps 3, 4, and 5 are all dominated by the global baseline Lasso complexity, the total computational complexity of the ARL algorithm simplifies to:

\begin{equation}
\mathcal{C}_{\text{ARL}} = \mathcal{C}_{\text{base}}(n, p) + O(p^2) + O(G \cdot m \cdot 2^{g})
\end{equation}

For the special case of coordinate descent Lasso, this becomes:

\begin{equation}
\mathcal{C}_{\text{ARL}} = O(npk) + O(p^2) + O(G \cdot m \cdot 2^{g})
\end{equation}

\textbf{Apriori-based Optimization.} We optimize the co-occurrence analysis through a unified frequent itemset mining approach based on the Apriori algorithm. The key insight is that if a feature combination of size $k$ does not exceed the co-occurrence threshold $\tau_{\text{co}}$, then any larger combination containing it will also be infrequent.

Our algorithm proceeds iteratively from size-1 to size-$k$ feature combinations. At each level $k$, we generate candidate combinations of size $k$ exclusively from frequent combinations of size $k-1$, and prune candidates whose $(k-1)$-subsets are infrequent. This layer-by-layer approach ensures that infrequent combinations at level $k$ prevent the generation of all their supersets at levels $k+1, k+2, \ldots$, maximizing pruning effectiveness. Formally, for a candidate combination $c$ of size $k$, we require:
\begin{equation}
\forall s \subset c, |s| = k-1 \Rightarrow \text{support}(s) > \tau_{\text{co}}
\end{equation}

This approach reduces the search space from $O(2^g)$ to $O(|\mathcal{F}_1| + |\mathcal{F}_2| + \ldots + |\mathcal{F}_k|)$ for a problem group of size $g$, where $|\mathcal{F}_i|$ denotes the number of frequent $i$-combinations. In practice, this achieves pruning ratios of 80-95\%, with higher effectiveness as $\tau_{\text{co}}$ increases, while maintaining identical rescue results to exhaustive enumeration.

\textbf{Comparison with Existing Methods.} We compare ARL's complexity with two prominent ensemble-based feature selection methods:

Random Lasso performs $B$ bootstrap iterations, each running the baseline Lasso on $n$ samples with $q$ randomly selected features. Its complexity is:
\begin{equation}
\mathcal{C}_{\text{RLasso}} = B \cdot \mathcal{C}_{\text{base}}(n, q)
\end{equation}
where $q$ is the number of randomly selected features per iteration.

Stability Selection performs $B$ subsampling iterations, each running the baseline Lasso on $n/2$ samples with all $p$ features:
\begin{equation}
\mathcal{C}_{\text{SS}} = B \cdot \mathcal{C}_{\text{base}}(n/2, p)
\end{equation}

\textbf{Comparative Analysis.} Ignoring the co-occurrence analysis term, ARL's complexity is approximately:
\begin{equation}
\mathcal{C}_{\text{ARL}}^{*} = \mathcal{C}_{\text{base}}(n, p) + O(p^2)
\end{equation}

Both Random Lasso and Stability Selection have complexity proportional to $B \cdot \mathcal{C}_{\text{base}}$, where $B$ is typically 100-1000. The key differentiator is the co-occurrence analysis term $O(G \cdot m \cdot 2^{g'})$. ARL is more efficient when:

\begin{equation}
G \cdot m \cdot 2^{g} \ll B \cdot \mathcal{C}_{\text{base}}(n, p)
\end{equation}

The Apriori optimization substantially reduces the computational complexity from the naive $O(2^g)$ enumeration. In practice, this enables efficient processing of problem groups with sizes up to $g \leq 12-15$, which is readily achievable through correlation-based problem group identification for moderate-dimensional problems ($p \leq 10^4$).

Conversely, when problem groups are exceptionally large and the optimization cannot sufficiently reduce computational complexity, Random Lasso and Stability Selection may be more efficient. However, in our simulation experiments, our method is consistently faster than these two ensemble methods.

\textbf{Additional Optimization for Large Problem Groups.} We also provide an optional optimization measure for cases where problem groups become exceptionally large. In the problem group identification step, we can impose an additional constraint requiring that the correlation coefficient between any two features within a problem group exceeds the threshold. This transforms the problem of group identification from finding connected components in a correlation graph to finding maximal cliques, ensuring that every pair of features within a group satisfies the correlation requirement.

While this pairwise constraint may be less precise than the transitive grouping approach in capturing complex correlation structures for Lasso's combinatorial optimization, it provides a significant computational advantage by bounding the maximum group size. Specifically, if we denote the maximum clique size as $g_{\max}$, then the co-occurrence analysis complexity becomes $O(G \cdot m \cdot 2^{g_{\max}})$, where $g_{\max}$ is typically much smaller than the average group size $g$ in the transitive approach.

This trade-off between correlation structure precision and computational efficiency allows practitioners to adapt the algorithm to their specific computational constraints and problem characteristics.

While increasing the correlation threshold can improve computational efficiency, this introduces a trade-off. The number of potentially rescuable features is related to the problem group size, $g$; a higher threshold leads to smaller groups, thus reducing the number of rescue candidates within each group. The ideal scenario for ARL is the presence of many distinct but small problem groups (a large $G$ and a small $g$). This configuration maximizes the total number of features that can be considered for rescue across the entire dataset, while ensuring the computational cost for analyzing each group remains manageable.

\subsection{Theorem}

Our proposed algorithm is supported by theoretical guarantees demonstrating its ability to distinguish between TR and FR feature groups. The full proofs are provided in the Appendix. The theoretical framework rests on the following assumptions.

\textbf{Assumption 1.1 (Correlation Structure).}
Let $G \subseteq \{1, ..., p\}$ be a problem group of features. We distinguish between two types of groups based on their true, underlying population correlation structure.

\begin{itemize}
    \item If G is a \textbf{TR group}, its population correlation matrix $\Sigma_{G}$ has off-diagonal entries $(\Sigma_{G})_{ij}=\rho$, where $\rho \in (1-\epsilon, 1)$ for a small constant $\epsilon > 0$. These features are inherently and systematically highly collinear.
    \item If G is a \textbf{Falsely Redundant (FR) group}, its population correlation matrix $\Sigma_{G}$ has off-diagonal entries $(\Sigma_{G})_{ij} = \rho_{ij}$ where $0 \le \rho_{ij} \le \rho_{max}$, for a constant $\rho_{max}$ that is significantly smaller than 1 (e.g., $\rho_{max} < 0.5$).
\end{itemize}

\textbf{Assumption 1.2 (Comparable Contribution).}
For any TR group $G_{TR}$ and FR group $G_{FR}$ of approximately the same size, the sum of their true coefficients is comparable. Specifically, there exists a constant $c \ge 1$ such that:
$$ \frac{1}{c} \sum_{k \in G_{FR}}\beta_{k}^{*} \le \sum_{j \in G_{TR}}\beta_{j}^{*} \le c \sum_{k \in G_{FR}}\beta_{k}^{*} $$
Furthermore, all coefficients $\beta_{j}^{*}$ within these groups are assumed to be non-zero.

\textbf{Assumption 1.3 (Lasso Selection Probability).}

It is a well-established property of the Lasso estimator, supported by extensive empirical and theoretical studies \cite{14,15}, that its variable selection behavior is highly sensitive to multicollinearity. In the presence of a group of extremely correlated features (a TR group), the L1 penalty induces a strong competition, making the selection of a single representative feature far more likely than the selection of the entire group. Conversely, for a group of features that are only moderately correlated and share a comparable collective contribution (an FR group), each feature retains a degree of unique information, making a joint selection more probable. We formalize this differential behavior as a foundational assumption for our analysis. Let $\hat{S}$ be the support of the Lasso solution. For any candidate set $C$ of size $|C| \ge 2$:
$$ P(C \subseteq \hat{S})_{\text{FR}} > P(C \subseteq \hat{S})_{\text{TR}} $$
For notational convenience, we denote these probabilities as $p_{C,\text{FR}}$ and $p_{C,\text{TR}}$ respectively, where $p_{C,\text{FR}}$ and $p_{C,\text{TR}}$ are distinct constants in the interval $(0, 1)$.

\textbf{Assumption 1.4 (Bounded Covariance).}
Let the random variable $X_{d}=\mathbb{I}(C\subseteq S_{d})$ indicate whether a candidate set C is fully selected in subset d. The sum of covariances of these variables satisfies the sub-quadratic growth condition:
$$\sum_{i \ne j}\text{Cov}(X_{i},X_{j})=o(m^{2})$$

Based on these assumptions, we establish our main theorem.

\textbf{Theorem 1.1 (Separability of Problem Groups).}
For any candidate set $C \subseteq G$ (where $|C| \ge 2$), the expected co-occurrence frequency is strictly higher for an FR group than for a TR group, i.e., $E[\text{count}(C)_{\text{FR}}] > E[\text{count}(C)_{\text{TR}}]$. Consequently, for a sufficiently large number of subsets $m$, there exists an integer threshold $\tau_{co}$ that can separate the two types of problem groups with high probability.

\section{Simulation studies}

Our algorithm is predicated on the existence of problem groups; in their absence, it degenerates to the Lasso baseline. Consequently, our simulation studies focus solely on scenarios in which our method is applicable. The first section will introduce our simulation setup. The second section will present an ablation study of our method benchmarked against nonensemble approaches. In the third section, we will compare ARL with existing mainstream ensemble methods and conduct the corresponding ablation studies.

The hyperparameter settings for the simulation studies are detailed in Appendix B.1.
\subsection{Simulation Setup}

\noindent\textbf{General Simulation Setup.} In our simulation study, the data are generated from a linear model $y = X\beta + \epsilon$, where $X \in \mathbb{R}^{N \times p}$ is the feature matrix, with $N$ being the sample size and $p$ the total number of features. The term $\epsilon \sim \mathcal{N}(0, \sigma^2)$ represents Gaussian white noise. The Signal-to-Noise Ratio (SNR) is defined as:
$$
\text{SNR} = \frac{\text{Var}(X\beta)}{\sigma^2}
$$
The feature matrix $X$ comprises predictors organized into four distinct categories to emulate complex real-world scenarios:
\begin{itemize}
    \item \textbf{Determined Important (DI) features:} These features are drawn independently from a standard normal distribution $\mathcal{N}(0,1)$.
    \item \textbf{TR features:} Each group of features is generated via a latent variable model $X_j = \rho L + \sqrt{1-\rho^2} E_j$ with a shared component $L \sim \mathcal{N}(0,1)$ and idiosyncratic errors $E_j \sim \mathcal{N}(0,1)$, which ensures that these features have real intra-group correlation.
    \item \textbf{Falsely Redundant (FR) features:} These are constructed by first generating base variables from $\mathcal{N}(0,1)$ and then adding a sample-specific random shock $k_i \cdot c$ to all features within a group. The correlation of this set of features is caused by a mean shift, and there is no real correlation itself.
    \item \textbf{Unimportant (U) features:} These are also drawn independently from $\mathcal{N}(0,1)$, and their true coefficients are strictly zero.
\end{itemize}

\noindent\textbf{EXAMPLE 1.} We instantiate the general setup with the following specific parameters. The feature matrix contains $p=1000$ predictors, comprising: 100 DI features; 100 TR features, organized into 10 groups of 10; 100 FR features, also organized into 10 groups of 10; and 700 U features. For the TR groups, the intra-group correlation is set to $\rho=0.95$. For the FR groups, the magnitude of the shift is $c=5.0$. To ensure fairness, the true coefficients $\beta_j$ for the three informative groups (DI, TR, and FR) are drawn randomly from a uniform distribution $U[0.1, 10]$.

Our experiments are designed to assess performance across multiple dimensions. First, in a baseline setting of $N=1000$, we evaluate performance across low, medium, and high Signal-to-Noise Ratios (SNR = 1, 3, and 6). Second, to examine the impact of dimensionality, we fix SNR at 3 and test two distinct scenarios relative to the feature dimension ($p=1000$): a classical setting with more samples than features ($N=2000$) and a challenging high-dimensional setting with fewer samples than features ($N=500$).

For the subsequent experiments (Examples 2, 3, and 4), we adopt a consistent high-dimensional setting. This setting is defined by a \textbf{sample size of $N=1000$}, a \textbf{feature dimension of $p=1000$}, and a \textbf{fixed Signal-to-Noise Ratio of $\text{SNR}=3$}.

\noindent\textbf{EXAMPLE 2.} Building on the general setup, we designed Example 2 to evaluate the algorithm's performance under a more severe setting. In Example 1, while the TR and FR groups were both highly correlated internally, the two types of groups were independent of each other. The core challenge of Example 2 is to break this independence by mixing TR and FR features within the same group and inducing a strong correlation between them, testing whether an algorithm can still accurately distinguish their fundamental differences.

Each MR group is made up of both TR and FR variables. We apply a unified, sample-driven, stepwise mean shift to the entire MR group. This operation not only creates a high macroscopic correlation among all variables within the group, but, more critically, it establishes a strong link between the previously uncorrelated TR and FR variables. This forces an algorithm to analyze a seemingly homogeneous, highly correlated cluster and partition it into two subsets of entirely different origins and properties, thus constituting a more advanced challenge.

To this end, we introduce a \textbf{Mixed Redundancy (MR) group}. The feature matrix contains 100 DI features, 10 MR groups, and 700 U features. Each MR group contains 10 TR variables and 10 FR variables, resulting in a total of 200 MR features. The true coefficients $\beta_j$ for all informative features (all features within the DI and MR groups) are randomly drawn from a uniform distribution $U[0.1, 10]$.

\noindent\textbf{EXAMPLE 3.} To evaluate the robustness of the ARL algorithm under different feature configurations, this example introduces three types of modifications to the feature composition, based on the setup of Example 1.
\begin{enumerate}
    \item\textbf{Setting1:} The 700 unimportant features are divided into 350 Correlated Unimportant ($U_{\text{corr}}$) features and 350 Purely Unimportant ($U_{\text{pure}}$) features. The $U_{\text{corr}}$ features are generated via the model $X_k = \rho_{\text{noise}} \text{PC}1 + \sqrt{1-\rho_{\text{noise}}^2} E_k$, with $\rho_{\text{noise}}=0.5$, where $\text{PC}1$ is the first principal component of the 300 informative features. This setting is designed to test the algorithm's performance when noise variables themselves exhibit correlation.
    
    \item\textbf{Setting2:} The number of DI features is increased to 300, reducing the unimportant features to 500. This scenario assesses performance when easily distinguishable features (DI) are more prevalent than features in complex correlation groups (TR and FR).
    
    \item\textbf{Setting3:} The quantity of all informative categories is doubled (200 DI, 200 TR, and 200 FR), decreasing the noise features to 400. This configuration evaluates the algorithm's behavior in a high-signal regime where the number of significant features exceeds that of noise features.
\end{enumerate}

\noindent\textbf{EXAMPLE 4.} To test the stability of the algorithm with coefficients of varying signs, this scenario is based on the high-dimensional setting. The key modification is that the true coefficients for all informative features are now drawn from a uniform distribution $U[-10, 10]$.

\noindent\textbf{Experimental Repetition and Reporting.} Finally, each experimental configuration is repeated on 100 independently generated datasets. Because the coefficients $\beta_j$ for all informative features are resampled from their distribution for each dataset, this procedure rigorously validates the robustness of our proposed method across various combinations of coefficient values. The results reported in this paper are the averages over these 100 trials.

\noindent\textbf{Evaluation Metrics.}
We chose precision, recall, and F1 score as our evaluation metrics. Since features within a TR group are generated with a true correlation of 0.95, we consider it optimal to select exactly one feature under such high collinearity. For a FR group, whose features are independent, all features must be selected.

Based on this, the counts for True Positives (TP), False Positives (FP), and False Negatives (FN) are defined as follows:

\begin{itemize}
    \item For each \textbf{TR group}:
    \begin{itemize}
        \item Selecting exactly one feature contributes to one TP.
        \item Selecting more than one feature contributes one TP for the first feature and one FP for each additional feature.
        \item Failing to select any feature contributes one FN.
    \end{itemize}
    \item For each \textbf{FR group}:
    \begin{itemize}
        \item Each correctly selected feature contributes one TP.
        \item Each feature that is not selected contributes one FN.
    \end{itemize}
\end{itemize}

The evaluation metrics are calculated based on the total counts aggregated across all groups as follows:

$$
\text{Precision} = \frac{\text{TP}}{\text{TP} + \text{FP}}
$$

$$
\text{Recall} = \frac{\text{TP}}{\text{TP} + \text{FN}}
$$

$$
\text{F1-score} = 2 \cdot \frac{\text{Precision} \cdot \text{Recall}}{\text{Precision} + \text{Recall}}
$$
We also present the total number of features selected by all methods in the simulation experiments in the appendix.

\noindent\textbf{BASELINE METHODS.} In the simulation experiments of this paper, Adaptive Lasso (ALasso), CVLasso, and Elastic Net (Enet) are selected as baseline Lasso methods. Ablation experiments are conducted using ARL. Furthermore, ARL is compared with two other ensemble methods, Stability Selection (SS) and Random Lasso (RLasso), for which ablation experiments are also performed. Furthermore, the hyperparameters for all methods involved in the simulation experiments of this paper are determined using cross-validation (CV) or common settings; detailed configurations will be presented in the appendix.

\subsection{Ablation Study against Non-Ensemble Methods}

\begin{table*}[h!]
\centering
\caption{The table displays the results for the baseline lasso methods and their ARL-enhanced versions across the four examples. The metrics are presented as $\substack{\text{F1-Score} \\ \text{(Precision / Recall)}}$, and for all metrics, higher is better.}
\label{Table 1}
\begin{tabular}{@{}llcccccc@{}}
\toprule
& & CVLasso & ARL-Lasso & ALasso & ARL-ALasso & Enet & ARL-Enet \\ \midrule
\multirow{5}{*}{\textbf{Example 1}} & \multicolumn{7}{l}{\textit{N = 1000}} \\ \addlinespace
& SNR=1 & $\substack{0.214 \\ (52.39/13.67)}$ & $\substack{0.272 \\ (51.75/19.85)}$ & $\substack{0.230 \\ (26.86/20.52)}$ & $\substack{0.344 \\ (34.70/34.92)}$ & $\substack{0.351 \\ (21.37/98.50)}$ & $\substack{0.349 \\ (21.22/98.50)}$ \\
& SNR=3 & $\substack{0.354 \\ (58.44/25.80)}$ & $\substack{0.495 \\ (48.40/51.58)}$ & $\substack{0.334 \\ (34.81/32.93)}$ & $\substack{0.478 \\ (42.67/55.21)}$ & $\substack{0.353 \\ (21.48/98.15)}$ & $\substack{0.350 \\ (21.30/98.15)}$ \\
& SNR=6 & $\substack{0.426 \\ (53.98/35.50)}$ & $\substack{0.486 \\ (46.61/51.67)}$ & $\substack{0.388 \\ (36.07/42.59)}$ & $\substack{0.483 \\ (42.29/56.96)}$ & $\substack{0.354 \\ (21.68/96.95)}$ & $\substack{0.352 \\ (21.50/96.95)}$ \\ \addlinespace
& \multicolumn{7}{l}{\textit{SNR = 3}} \\ \addlinespace
& N=500 & $\substack{0.303 \\ (54.09/21.42)}$ & $\substack{0.470 \\ (51.09/45.63)}$ & $\substack{0.307 \\ (29.97/33.30)}$ & $\substack{0.435 \\ (39.04/51.07)}$ & $\substack{0.352 \\ (21.44/98.20)}$ & $\substack{0.350 \\ (21.27/98.20)}$ \\
& N=2000 & $\substack{0.399 \\ (57.42/30.89)}$ & $\substack{0.512 \\ (49.08/54.11)}$ & $\substack{0.334 \\ (34.81/32.93)}$ & $\substack{0.478 \\ (42.67/55.21)}$ & $\substack{0.353 \\ (21.53/97.88)}$ & $\substack{0.351 \\ (21.36/97.88)}$ \\ \midrule
\textbf{Example 2} &  & $\substack{0.278 \\ (46.64/20.06)}$ & $\substack{0.534 \\ (55.24/52.30)}$ & $\substack{0.271 \\ (28.92/26.19)}$ & $\substack{0.423 \\ (40.22/45.79)}$ & $\substack{0.349 \\ (21.23/98.20)}$ & $\substack{0.348 \\ (21.23/98.20)}$ \\ \midrule
\multirow{3}{*}{\textbf{Example 3}} & Setting1 & $\substack{0.372 \\ (68.12/25.77)}$ & $\substack{0.514 \\ (52.09/51.76)}$ & $\substack{0.390 \\ (50.19/32.53)}$ & $\substack{0.551 \\ (55.00/55.68)}$ & $\substack{0.493 \\ (32.95/98.10)}$ & $\substack{0.489 \\ (32.59/98.10)}$ \\
& Setting2 & $\substack{0.259 \\ (68.90/16.10)}$ & $\substack{0.360 \\ (53.49/27.52)}$ & $\substack{0.350 \\ (53.97/26.32)}$ & $\substack{0.446 \\ (57.29/36.82)}$ & $\substack{0.581 \\ (41.44/96.99)}$ & $\substack{0.577 \\ (41.10/96.99)}$ \\
& Setting3 & $\substack{0.368 \\ (69.95/25.08)}$ & $\substack{0.493 \\ (56.54/46.25)}$ & $\substack{0.386 \\ (60.47/28.68)}$ & $\substack{0.566 \\ (63.25/51.40)}$ & $\substack{0.591 \\ (42.80/98.64)}$ & $\substack{0.591 \\ (42.18/98.64)}$ \\ \midrule
\textbf{Example 4} & & $\substack{0.311 \\ (55.89/22.00)}$ & $\substack{0.496 \\ (48.57/51.73)}$ & $\substack{0.310 \\ (32.23/30.91)}$ & $\substack{0.427 \\ (38.73/48.51)}$ & $\substack{0.352 \\ (22.35/95.14)}$ & $\substack{0.357 \\ (21.96/95.85)}$ \\ \bottomrule
\end{tabular}
\end{table*}

As shown by the results in Table \ref{Table 1}, when using ALasso and CVLasso as baseline methods, ARL achieves a substantial improvement in F1-score across different settings of SNR and N, indicating a better simultaneous control over both types of Lasso errors. For ALasso, ARL manages to improve both precision and recall. However, when Enet is used as the baseline, the performance is unsatisfactory. This is determined by the inherent bias of different baseline methods. Enet is biased towards selecting more features, which diminishes the recovery space for ARL. In our experiments, we observed that Enet selected nearly all FR features in the initial global selection, leaving no room for ARL to perform rescues. This is also related to our simulation configuration of 700 noise variables versus 300 signal variables. While this leads to a decent recall for Enet, its precision is low compared to other methods. Therefore, methods like Enet, which are inherently inclined to retain more features, are not suitable baselines for ARL.

From the results of example 2, it is evident that the co-existence of true and false redundancy poses a significant challenge for the baseline methods, with nearly all of them experiencing a drastic drop in F1-score in this difficult setting. Nevertheless, ARL's performance remains exceptional even in this scenario. Due to the extremely low discriminative power of the baseline lasso in this environment, the magnitude of the F1 score improvement brought by ARL is even greater than that under the identical setting in Example 1.

The results from Example 3 indicate that ARL remains stable in a more complex environment, delivering an F1-score improvement similar to that observed in Example 1. This demonstrates that ARL is well-suited for complex environments and exhibits excellent robustness.

Regarding the hyperparameters $m$ and $\tau_{co}$, we determined them empirically in our simulations. A heuristic guideline is that with fewer samples, we should partition the data into fewer subsets to ensure that each subset is large enough for the Lasso results to be meaningful. For higher SNR, the increased information content leads baseline methods to generally select more features; this requires us to raise the threshold $\tau_{co}$ to prevent the inclusion of too many truly redundant features. In our experiments, for N = 2000, 1000, and 500, we chose $m$ to be 30, 30, and 10, respectively. For SNR = 1, 3, and 6, we chose $\tau_{co}$ to be 1, 1, and 2, respectively. These are purely empirical choices without any tuning; in practice, they can be determined via cross-validation.

\begin{table*}[h!]
\centering
\caption{Average number of features selected by each algorithm over 100 random datasets}
\label{Table 2}
\begin{tabular}{@{}llcccccc@{}}
\toprule
& & CVLasso & ARL-Lasso & ALasso & ARL-ALasso & Enet & ARL-Enet \\ \midrule
\multirow{5}{*}{\textbf{Example 1}} & \multicolumn{7}{l}{\textit{N = 1000}} \\ \addlinespace
& SNR=1 & 57.4 & 84.3 & 164.0 & 214.6 & 968.0 & 974.6 \\
& SNR=3 & 95.8 & 226.8 & 205.2 & 277.4 & 959.4 & 967.5 \\
& SNR=6 & 140.6 & 235.7 & 252.4 & 287.2 & 939.0 & 946.9 \\ \addlinespace
& \multicolumn{7}{l}{\textit{SNR = 3}} \\ \addlinespace
& N=500 & 85.9 & 194.5 & 244.9 & 286.6 & 961.8 & 969.5 \\
& N=2000 & 115.3 & 235.1 & 205.2 & 277.4 & 954.7 & 962.4 \\ \midrule
\textbf{Example 2} &  & 92.8 & 201.0 & 195.5 & 243.5 & 971.4 & 971.4 \\ \midrule
\multirow{3}{*}{\textbf{Example 3}} & Setting1 & 102.8 & 233.4 & 198.0 & 273.5 & 959.2 & 966.0 \\
& Setting2 & 97.7 & 214.0 & 202.9 & 266.2 & 959.4 & 967.5 \\
& Setting3 & 152.0 & 357.4 & 201.0 & 342.7 & 968.1 & 982.3 \\ \midrule
\textbf{Example 4} & & 86.2 & 225.5 & 208.5 & 267.7 & 894.3 & 916.9 \\ \bottomrule
\end{tabular}
\end{table*}
The results in Table \ref{Table 2} suggest that the baseline for ARL should be an algorithm that favors sparse solutions, as this provides ARL with greater scope for improvement.

\subsection{Comparison with and Ablation Study on Mainstream Ensemble Methods}

\begin{table*}[h!]
\centering
\caption{The table displays the results for the ensemble lasso methods and their ARL-enhanced versions across the four examples. The metrics are presented as $\substack{\text{F1-Score} \\ \text{(Precision / Recall)}}$, and for all metrics, higher is better.}
\label{Table 3}
\begin{tabular}{@{}llcccc@{}}
\toprule
& & SS & ARL-SS & R.Lasso & ARL-RLasso \\ \midrule
\multirow{5}{*}{\textbf{Example 1}} & \multicolumn{5}{l}{\textit{N = 1000}} \\ \addlinespace
& SNR=1 & $\substack{0.172 \\ (28.17/12.43)}$ & $\substack{0.173 \\ (28.29/12.50)}$ & $\substack{0.430 \\ (31.67/67.81)}$ & $\substack{0.408 \\ (28.79/70.61)}$ \\
& SNR=3 & $\substack{0.208 \\ (44.10/13.67)}$ & $\substack{0.253 \\ (49.58/17.04)}$ & $\substack{0.407 \\ (28.04/74.95)}$ & $\substack{0.395 \\ (26.80/75.76)}$ \\
& SNR=6 & $\substack{0.232 \\ (57.50/14.55)}$ & $\substack{0.304 \\ (64.87/19.91)}$ & $\substack{0.387 \\ (25.34/82.03)}$ & $\substack{0.383 \\ (25.05/82.09)}$ \\ \addlinespace
& \multicolumn{5}{l}{\textit{SNR = 3}} \\ \addlinespace
& N=500 & $\substack{0.195 \\ (76.75/11.21)}$ & $\substack{0.243 \\ (80.75/14.34)}$ & $\substack{0.405 \\ (27.59/77.20)}$ & $\substack{0.394 \\ (26.44/78.24)}$ \\
& N=2000 & $\substack{0.266 \\ (17.42/56.68)}$ & $\substack{0.272 \\ (17.82/58.17)}$ & $\substack{0.409 \\ (28.09/75.73)}$ & $\substack{0.403 \\ (27.45/76.11)}$ \\ \midrule
\textbf{Example 2} & & $\substack{0.215 \\ (26.43/18.27)}$ & $\substack{0.256 \\ (30.47/22.17)}$ & $\substack{0.428 \\ (32.33/63.99)}$ & $\substack{0.435 \\ (32.54/66.34)}$ \\ \midrule
\multirow{3}{*}{\textbf{Example 3}} & Setting1 & $\substack{0.237 \\ (56.66/15.02)}$ & $\substack{0.290 \\ (61.29/18.20)}$ & $\substack{0.523 \\ (40.23/75.31)}$ & $\substack{0.506 \\ (38.00/75.89)}$ \\
& Setting2 & $\substack{0.172 \\ (61.16/10.03)}$ & $\substack{0.188 \\ (63.53/11.08)}$ & $\substack{0.531 \\ (45.88/63.52)}$ & $\substack{0.518 \\ (43.75/63.88)}$ \\
& Setting3 & $\substack{0.332 \\ (49.07/25.29)}$ & $\substack{0.332 \\ (49.09/25.30)}$ & $\substack{0.614 \\ (49.88/79.99)}$ & $\substack{0.577 \\ (45.16/80.18)}$ \\ \midrule
\textbf{Example 4} & & $\substack{0.157 \\ (56.43/9.19)}$ & $\substack{0.181 \\ (60.11/10.70)}$ & $\substack{0.404 \\ (27.53/76.83)}$ & $\substack{0.395 \\ (26.41/78.79)}$ \\ \bottomrule
\end{tabular}
\end{table*}

\begin{table*}[h!]
\centering
\caption{Average number of features selected by each algorithm over 100 random datasets}
\label{tab:ensemble_feature_counts}
\begin{tabular}{@{}llcccc@{}}
\toprule
& & SS & ARL-SS & R.Lasso & ARL-RLasso \\ \midrule
\multirow{5}{*}{\textbf{Example 1}} & \multicolumn{5}{l}{\textit{N = 1000}} \\ \addlinespace
& SNR=1 & 93.1 & 93.2 & 454.1 & 518.2 \\
& SNR=3 & 65.6 & 72.7 & 566.6 & 597.1 \\
& SNR=6 & 53.5 & 64.8 & 682.2 & 690.5 \\ \addlinespace
& \multicolumn{5}{l}{\textit{SNR = 3}} \\ \addlinespace
& N=500 & 30.9 & 37.4 & 593.1 & 625.1 \\
& N=2000 & 681.8 & 684.9 & 568.9 & 584.5 \\ \midrule
\textbf{Example 2} & & 145.1 & 153.3 & 419.7 & 431.8 \\ \midrule
\multirow{3}{*}{\textbf{Example 3}} & Setting1 & 75.8 & 82.4 & 597.0 & 622.4 \\
& Setting2 & 67.5 & 71.8 & 570.1 & 599.8 \\
& Setting3 & 217.4 & 217.5 & 675.2 & 746.2 \\ \midrule
\textbf{Example 4} & & 34.7 & 37.9 & 590.8 & 629.6 \\ \bottomrule
\end{tabular}
\end{table*}

The results in Table \ref{Table 3} show that SS (based on ALasso) and RLasso (based on CVLasso) exhibit highly unstable performance, a common drawback of existing ensemble approaches. In contrast, ARL, by leveraging the advantages of the A-R framework, achieves a more controllable ensemble. It not only realizes a general improvement in F1-score over the baseline methods but also, in most scenarios, shows a greater margin of improvement than the other two ensemble methods.

Furthermore, ARL can itself use SS and RLasso as baseline methods. It yields comprehensive improvements in both precision and recall when applied to SS. However, it is not suitable for RLasso for reasons similar to Enet. RLasso also tends to retain more features, a characteristic that makes it an unsuitable baseline for ARL and likewise results in low precision.

\section{Real Data}

To evaluate the practical effectiveness of our proposed ARL algorithm, we conduct comprehensive experiments on real-world datasets with inherent complexity and noise characteristics absent in controlled simulations.

The hyperparameter settings for the real data experiment are detailed in Appendix B.2.

\subsection{Ames Housing}

\textbf{Dataset Description:} We employ the Ames Housing dataset, a comprehensive real estate dataset containing 1,460 residential properties in Ames, Iowa, with 80 explanatory variables describing various aspects of residential homes \cite{16}.

This dataset was chosen because it represents a classic regression problem with a complex mixture of numerical and categorical variables. This structure poses a significant challenge for feature selection, as predictive performance is not necessarily improved by simply selecting more or fewer features. Furthermore, the presence of missing values, combined with the one-hot encoding required for categorical variables, gives rise to numerous non-informative yet potentially correlated features, creating a scenario analogous to the TR groups defined in our study.

The prediction target is the logarithm of sale price, chosen to stabilize variance and improve model interpretability.

\textbf{Data Preprocessing:} Missing Value Treatment: We implement a systematic approach to handle missing data based on variable type. For numerical features, missing values are imputed using the median value of the respective feature, providing robustness against outliers. For categorical features, missing values are treated as a separate category labeled "Missing", recognizing that missingness itself may carry predictive information.

Categorical Variable Encoding: Categorical variables were transformed using one-hot encoding, which generated 261 binary features.

Feature Scaling: All features are standardized using z-score normalization to ensure equal contribution scales across variables with different units and ranges. The standardization is performed using training set statistics and applied consistently to both training and test sets to prevent data leakage.

\textbf{Evaluation Metrics and Experimental Setup:} We randomly split the dataset into 80\% training and 20\% test sets. Model performance is evaluated using Root Mean Square Error (RMSE) and its normalized version:

\begin{equation}
\text{RMSE}_{\text{normalized}} = \frac{1}{\sigma_y} \sqrt{\frac{1}{n} \sum_{i=1}^{n} (y_i - \hat{y}_i)^2}
\end{equation}

Where $y_i$ represents the true log sale price, $\hat{y}_i$ denotes the predicted log sale price, $n$ is the number of test samples, and $\sigma_y$ is the standard deviation of the target variable in the training set. For each method, we first perform an Ordinary Least Squares (OLS) regression on the test set using the features selected by the respective algorithm. The predictive performance is then evaluated using the Normalized Root Mean Square Error ($\text{RMSE}_{\text{normalized}}$), which expresses the prediction error as a fraction of the target variable's natural variability, with values closer to zero indicating better performance.

\begin{table}[h!]
\centering
\caption{Performance Comparison of Feature Selection Methods on the Ames Housing Dataset. The performance is evaluated using Normalized RMSE.}
\label{tab:ames_performance}
\begin{tabular}{lcc}
\toprule
\textbf{Method} & \textbf{No. of features selected} & \textbf{RMSE} \\
\midrule
Enet        & 258 & 1.2748 \\
ALasso     & 88  & 0.3614 \\
SS          & 218 & 0.3587 \\
RLasso     & 170 & 0.3589 \\
CVLasso     & 54  & 0.3575 \\
ARL-Lasso   & 59  & 0.3571 \\
\bottomrule
\end{tabular}
\end{table}

As shown in Table \ref{tab:ames_performance}, Enet selects nearly all features, and its predictive performance is even weaker than that of a mean-only prediction. SS selects the second-largest number of features, but its RMSE is only surpassed by CVLasso and ARL-Lasso. Although CVLasso uses the fewest features, its RMSE is superior to both Enet and SS. The ARL algorithm, building upon CVLasso, finds five additional features that successfully reduce the RMSE, demonstrating its effectiveness.

\subsection{Breast Cancer}
We employ the Breast Cancer Wisconsin (Diagnostic) dataset \cite{17}, a widely used benchmark for binary classification. The dataset contains 569 instances, each described by 30 numeric features computed from a digitized image of a fine needle aspirate (FNA) of a breast mass.

In a departure from the previously utilized economics dataset, we select a classic medical dataset for this benchmark. Distinct from the prior regression-based tasks, this is a classification problem. For each method, we first perform a logistic regression on the test set using the features selected by the respective algorithm. The predictive performance is then evaluated using Accuracy (ACC), defined as:
\begin{equation}
    \text{ACC} = \frac{1}{n} \sum_{i=1}^{n} \mathbb{I}(y_i = \hat{y}_i)
\end{equation}
where $n$ is the total number of samples, $y_i$ is the true label for sample $i$, $\hat{y}_i$ is the predicted label, and $\mathbb{I}(\cdot)$ is the indicator function.
Since the benchmark is relatively simple, in order to increase the difficulty, we randomly split the dataset into 70\% training and 30\% test sets.

\begin{table}[h!]
\centering
\caption{Performance Comparison on the Breast Cancer Wisconsin Dataset}
\label{tab:breast_cancer_performance}
\begin{tabular}{lcc}
\toprule
\textbf{Method} & \textbf{No. of features selected} & \textbf{ACC} \\
\midrule
SS          & 3  & 0.9474 \\
ALasso      & 14 & 0.9708 \\
RLasso      & 19 & 0.9766 \\
CVLasso     & 23 & 0.9825 \\
Enet        & 24 & 0.9825 \\
ARL-ALasso  & 17 & 0.9825 \\
ARL-Enet    & 26 & 0.9883 \\
\bottomrule
\end{tabular}
\end{table}

As shown by the results in Table \ref{tab:breast_cancer_performance}, the Enet-based ARL algorithm achieves the optimal predictive accuracy. Furthermore, the Alasso-based ARL algorithm secures the second-highest accuracy using only 17 features. This performance surpasses that of Enet, Alasso, and Rlasso, which selected a larger number of features, and strongly demonstrates the effectiveness of the ARL algorithm.

\newpage
\section{Discussion and Limitations}

Computational Complexity and Sample Size Requirements: The computational complexity of the co-occurrence analysis is exponential in the size of the problem groups under consideration. Although optimizations have been implemented to control group size, in scenarios with a very high feature dimension ($p$), if individual problem groups become exceptionally large, a preliminary dimensionality reduction step using other feature selection methods may be necessary. Furthermore, the total sample size ($N$) cannot be too small, as this would result in subsets with insufficient samples, thereby undermining the reliability of the subset Lasso results. Future work could explore the possibility of augmenting the overall sample size through techniques such as bootstrapping.

Scope of Application The ARL framework is fundamentally designed to strike a balance between mitigating multicollinearity (by consolidating TR groups) and preventing feature omission (by rescuing FR groups). Consequently, for applications with highly skewed objectives, ARL may not be the optimal choice. For instance, in scenarios where the cost of feature omission far outweighs any concern for model complexity, or conversely, where achieving extreme model parsimony is the sole priority, more specialized methods might be preferable.

Hyperparameter Sensitivity The ARL algorithm is sensitive to its hyperparameter settings, which require careful selection based on the specific problem context and analytical goals. The following heuristics can guide this process:
\begin{itemize}
    \item \textbf{Number of Subsets ($m$):} A larger total training sample allows for a greater number of subsets ($m$). A proportional relationship should be maintained between the total sample size and $m$ to ensure that each subset is large enough for reliable Lasso estimation.
    \item \textbf{Co-occurrence Threshold ($\tau_{co}$):} In settings with a higher signal-to-noise ratio or where precision is paramount, a higher co-occurrence threshold should be chosen to enforce a stricter standard for feature rescue.
    \item \textbf{Correlation Threshold ($\tau_{corr}$):} The choice of this threshold reflects the trade-off between concerns over multicollinearity and feature omission. If mitigating multicollinearity is the primary concern, a higher threshold is appropriate to form more tightly-knit problem groups. Conversely, if preventing feature omission is more critical, a lower threshold can be used to allow more potentially valuable features to be considered for rescue.
\end{itemize}
Future research can explore two related problems. The first is the dual to the issue addressed in this paper: using the same principles to identify and correct for spurious non-correlation, where features appear uncorrelated in the global dataset, but this relationship proves unstable in subsets. The second is the development of a unified framework that can simultaneously perform this correction for spurious non-correlation as well as the feature rescue from spurious correlation described herein.

\newpage

\bibliography{aoas-template}

\section*{Appendix}
\subsection*{A: Proofs of Theoretical Results}

\subsubsection*{A.1 Definition and Assumptions}

\textbf{Definition 1.1 (Lasso Objective Function).}
The objective function for the Lasso is given by:
$$J(\beta)=\frac{1}{2}||y-X\beta||_{2}^{2}+\lambda||\beta||_{1}$$

\textbf{Assumption 1.1 (Correlation Structure).}
Let $G \subseteq \{1, ..., p\}$ be a problem group of features. We distinguish between two types of groups based on their true, underlying population correlation structure.

\begin{itemize}
    \item If G is a \textbf{TR group}, its population correlation matrix $\Sigma_{G}$ has off-diagonal entries $(\Sigma_{G})_{ij}=\rho$, where $\rho \in (1-\epsilon, 1)$ for a small constant $\epsilon > 0$. These features are inherently and systematically highly collinear.
    \item If G is a \textbf{Falsely Redundant (FR) group}, its population correlation matrix $\Sigma_{G}$ has off-diagonal entries $(\Sigma_{G})_{ij} = \rho_{ij}$ where $0 \le \rho_{ij} \le \rho_{max}$, for a constant $\rho_{max}$ that is significantly smaller than 1 (e.g., $\rho_{max} < 0.5$).
\end{itemize}

\textbf{Assumption 1.2 (Comparable Contribution).}
For any TR group $G_{TR}$ and FR group $G_{FR}$ of approximately the same size, the sum of their true coefficients is comparable. Specifically, there exists a constant $c \ge 1$ such that:
$$ \frac{1}{c} \sum_{k \in G_{FR}}\beta_{k}^{*} \le \sum_{j \in G_{TR}}\beta_{j}^{*} \le c \sum_{k \in G_{FR}}\beta_{k}^{*} $$
Furthermore, all coefficients $\beta_{j}^{*}$ within these groups are assumed to be non-zero.

\textbf{Assumption 1.3 (Lasso Selection Probability).}

It is a well-established property of the Lasso estimator, supported by extensive empirical and theoretical studies \cite{14,15}, that its variable selection behavior is highly sensitive to multicollinearity. In the presence of a group of extremely correlated features (a TR group), the L1 penalty induces a strong competition, making the selection of a single representative feature far more likely than the selection of the entire group. Conversely, for a group of features that are only moderately correlated and share a comparable collective contribution (an FR group), each feature retains a degree of unique information, making a joint selection more probable. We formalize this differential behavior as a foundational assumption for our analysis. Let $\hat{S}$ be the support of the Lasso solution. For any candidate set $C$ of size $|C| \ge 2$:
$$ P(C \subseteq \hat{S})_{\text{FR}} > P(C \subseteq \hat{S})_{\text{TR}} $$
For notational convenience, we denote these probabilities as $p_{C,\text{FR}}$ and $p_{C,\text{TR}}$ respectively, where $p_{C,\text{FR}}$ and $p_{C,\text{TR}}$ are distinct constants in the interval $(0, 1)$.

\textbf{Assumption 1.4 (Bounded Covariance).}
Let the random variable $X_{d}=\mathbb{I}(C\subseteq S_{d})$ indicate whether a candidate set C is fully selected in subset d. The sum of covariances of these variables satisfies the sub-quadratic growth condition:
$$\sum_{i \ne j}\text{Cov}(X_{i},X_{j})=o(m^{2})$$

\subsubsection*{A.2 Main Theorem and Proof}

\textbf{Theorem 1.1 (Separability of Problem Groups).}

\textit{Statement:}
For any candidate set $C \subseteq G$ (where $|C| \ge 2$), the expected co-occurrence frequency is strictly higher for an FR group than for a TR group, i.e., $E[\text{count}(C)_{\text{FR}}] > E[\text{count}(C)_{\text{TR}}]$. Consequently, for a sufficiently large number of subsets $m$, there exists an integer threshold $\tau_{co}$ that can separate the two types of problem groups with high probability.

\textit{Proof.}
Let $\text{count}(C) = \sum_{d=1}^{m} \mathbb{I}(C \subseteq S_d)$ be the random variable for the co-occurrence frequency of a candidate set $C$ across $m$ subsets, where $\mathbb{I}(\cdot)$ is the indicator function and $S_d$ is the support of the Lasso solution on subset $d$.

First, we establish the inequality of the expected values. Let $p_C = P(C \subseteq S_d)$ be the probability of selection in any single subset. By the linearity of expectation:
$$ E[\text{count}(C)] = E\left[\sum_{d=1}^{m} \mathbb{I}(C \subseteq S_d)\right] = \sum_{d=1}^{m} E[\mathbb{I}(C \subseteq S_d)] = m \cdot p_C $$
From \textbf{Assumption 1.3}, we have the strict inequality $p_{C,\text{FR}} > p_{C,\text{TR}}$. Therefore, it directly follows that:
$$ E[\text{count}(C)_{\text{FR}}] = m \cdot p_{C,\text{FR}} > m \cdot p_{C,\text{TR}} = E[\text{count}(C)_{\text{TR}}] $$
This proves the first part of the theorem.

Next, we show that the observed frequency concentrates around its mean. The variance of the count is given by:
$$ \text{Var}(\text{count}(C)) = \sum_{d=1}^{m}\text{Var}(\mathbb{I}(C \subseteq S_d)) + \sum_{i\ne j}\text{Cov}(\mathbb{I}(C \subseteq S_i), \mathbb{I}(C \subseteq S_j)) $$
The first term, the sum of variances, is $m p_C(1-p_C) = O(m)$. From \textbf{Assumption 4}, the second term, the sum of covariances, exhibits sub-quadratic growth, i.e., $\sum_{i\ne j}\text{Cov}(\cdot) = o(m^2)$. Thus, the total variance is dominated by the covariance term for large $m$, giving $\text{Var}(\text{count}(C)) = o(m^2)$.

By applying Chebyshev's inequality to the sample mean frequency $\bar{p}_C = \text{count}(C)/m$, for any $\epsilon > 0$:
$$ P(|\bar{p}_C - p_C| \ge \epsilon) \le \frac{\text{Var}(\bar{p}_C)}{\epsilon^2} = \frac{\text{Var}(\text{count}(C))}{m^2 \epsilon^2} = \frac{o(m^2)}{m^2 \epsilon^2} $$
As $m \to \infty$, this probability converges to 0. This establishes that the sample frequency $\bar{p}_C$ converges in probability to its expected value $p_C$.

Since $\text{count}(C)_{\text{FR}}/m$ and $\text{count}(C)_{\text{TR}}/m$ converge in probability to the distinct constants $p_{C,\text{FR}}$ and $p_{C,\text{TR}}$ respectively, their distributions become increasingly concentrated around their different means. For any value $\tau'$ such that $p_{C,\text{TR}} < \tau' < p_{C,\text{FR}}$, the probability that the counts fall on the wrong side of the threshold $\tau_{co} = m\tau'$ vanishes as $m \to \infty$. Thus, for a sufficiently large $m$, a separating threshold exists with high probability.

\subsection*{B: Experimental Configuration}

\subsubsection*{B.1 Simulation studies}
To ensure fair comparison and reproducibility, we standardize all hyperparameter settings across baseline methods without additional tuning. This section details the specific configurations used for each algorithm in our experiments.

\textbf{Common Parameters.} All methods share the following universal settings: random seed is fixed at 42 for reproducibility, and convergence tolerance is set to 0.005 for global phases and 0.01 for subset phases to balance computational efficiency with solution quality.

\textbf{Cross-Validation Settings.} For methods requiring cross-validation, we employ 5-fold CV for global feature selection phases and 3-fold CV for subset analysis phases. This configuration provides reliable parameter estimation while maintaining computational tractability for subset operations.

\textbf{LassoCV Configuration.} The standard LassoCV method uses cross-validation to automatically determine the optimal regularization parameter $\alpha$ from sklearn's default candidate set, with no manual intervention in parameter selection.

\textbf{Adaptive Lasso Configuration.} For Adaptive Lasso, the ridge regression component uses 100 logarithmically spaced $\alpha$ candidates from $10^{-5}$ to $10^5$, with the optimal value selected via cross-validation. The subsequent Lasso step uses the globally determined $\alpha$ from LassoCV. The weight regularization parameter is set to $10^{-10}$ to ensure numerical stability.

\textbf{Ensemble Method Configurations.} Both Random Lasso and Stability Selection employ 200 bootstrap/subsampling iterations for global phases and 50 iterations for subset phases. Random Lasso uses bootstrap sampling (sampling with replacement of the original sample size), while Stability Selection uses subsampling without replacement to half the original sample size ($n/2$).

For Random Lasso, both feature sampling phases (q1 and q2) select 10\% of available features per iteration, with the baseline method being LassoCV for $\alpha$ determination.

For Stability Selection, the baseline method is ALasso with a critical modification: to avoid excessive computational burden, the ridge regression component uses a fixed $\alpha = 1.0$ rather than cross-validation selection. The selection threshold is set to 0.75, meaning features must be selected in at least 75\% of subsamples to be included in the final set. For large sample scenarios ($n = 2000$), where ridge regression with $\alpha = 1.0$ tends to select all features due to the non-sparse nature of ridge solutions, we substitute the global $\alpha$ determined by LassoCV to maintain meaningful feature selection.

\textbf{Elastic Net Configuration.} ElasticNetCV automatically determines both the $\alpha$ and $l1\_ratio$ parameters through cross-validation using sklearn's default parameter grids, ensuring optimal balance between L1 and L2 regularization.

\textbf{ARL-Specific Parameters.} For the ARL algorithm, we set the correlation threshold to 0.8 for problem group identification, the silhouette threshold to 0.5 for clustering basis evaluation, 50 final clusters for data partitioning, a minimum subset size of 20 samples, and a co-occurrence threshold of 1 for feature rescue.

This standardized configuration ensures that performance differences reflect algorithmic advantages rather than artifacts of parameter tuning, providing a fair basis for comparative evaluation.

\subsubsection*{B.2 Real Data}

The primary modification concerns the Stability Selection algorithm, where we replace the fixed ridge regression parameter with cross-validation selection. Specifically, the ridge regression component now uses RidgeCV with $\alpha$ candidates spanning $\text{logspace}(-5, 5, 100)$ and 3-fold cross-validation, automatically selecting the optimal regularization strength for each subsample. This enhancement improves the adaptive weight calculation and reduces sensitivity to the ridge parameter choice.

For the ARL algorithm \textbf{on the Ames Housing dataset}, we implement empirical parameter adjustments based on the dataset characteristics. Given training sets of approximately 1000 samples, we reduce the number of data subsets from $m = 50$ to $m = 30$ to ensure adequate sample size per subset while maintaining computational efficiency. Recognizing that real data exhibits less structured correlation patterns compared to controlled simulation data, we increase the co-occurrence threshold from $\tau_{\text{co}} = 1$ to $\tau_{\text{co}} = 3$ to enforce stricter consistency requirements for feature rescue. Additionally, we lower the correlation threshold for problem group identification from 0.8 to 0.75, allowing the algorithm to capture more subtle but meaningful correlation structures that may be present in real-world feature relationships.

These parameter adjustments reflect a more conservative approach suitable for the inherent noise and complexity of real datasets, ensuring that rescued features demonstrate robust co-occurrence patterns while maintaining the algorithm's ability to identify relevant correlation structures in practical applications.

For the ARL algorithm \textbf{on the Breast Cancer dataset}, based on the sample size of $N \approx 500$, we set the number of subsets to $m=10$, following the configuration used in our simulation experiments. All other parameters were kept consistent with the simulation setup.

\end{document}